Extremely short electromagnetic field pulse in supersymmetric electrodynamics

M.B. Belonenko, N.N. Konobeeva

Volgograd State University

mbelonenko@yandex.ru

We obtain the Maxwell`s equations used the supersymmetric action based on the actions for the scalar and spinor fields, which are built on the invariants of the electromagnetic field. We analyze the pulse instability in the framework of nonlinear electrodynamics without the approximation of slowly varying amplitudes and phases. We observe the collapse of an extremely short pulse. Within the framework of the Schwinger mechanism, the creation of scalar and spin particles is estimated.

**1. Introduction**

Historically, nonlinear electrodynamics (NE) appeared due to the fact that in classical electrodynamics the field of a point charge tends to infinity when approaching a point charge [1]. To get rid of this circumstance, many authors proposed nonlinear Lagrangians, which give the usual electrodynamic effects in the case of weak fields, but due to nonlinearity give a finite value of the field of a point charge at the origin, which coincides with the charge. Quite a lot of such Lagrangians can be proposed. And the choice of one or another of them is largely determined by the preferences of the authors. At present, the NE effects associated with the consideration of the Heisenberg-Euler Lagrangian [2, 3], which in the general case is given in the form of an integral, which depends on the scalar and pseudoscalar invariants of the electromagnetic field, are well studied. From the point of view of quantum electrodynamics, such nonlinearities arise due to the exchange of virtual electron-positron pairs between photons in a vacuum. Note also that this approach can be easily generalized to the case of "rapidly changing" fields, as described in [4, 5]. The Maxwell Lagrangian of the electromagnetic field is a first-order term and contains only one invariant (scalar).

Along with the Heisenberg-Euler Lagrangian, the study of the effects of nonlinear electrodynamics is possible within the framework of a different approach. So for the case of scalar and spinor electrodynamics, both the Heisenberg-Euler Lagrangian and the supersymmetric Lagrangian can be used. In this work, the Lagrangian proposed in [6] is chosen. We investigate the propagation of extremely short pulses in vacuum, which from the point of view of ordinary electrodynamics spread out due to dispersion. The question of the stability of such pulses in the case of NE is still open.

## 2. Basic equations

The supersymmetric Lagrangian proposed by M.J. Duff has the form [6, 7]:

$$L = L_{spin} + 2L_{scal},$$
$$L_{spin} = -\frac{1}{8\pi^2}\int_0^\infty \frac{dT}{T} e^{-m^2 T}\left(\frac{e^2 ab}{th(eaT) tg(ebT)} - \frac{e^2}{3}(a^2 - b^2) - \frac{1}{T^2}\right),$$
$$L_{scal} = \frac{1}{16\pi^2}\int_0^\infty \frac{dT}{T} e^{-m^2 T}\left(\frac{e^2 ab}{sh(eaT) sin(ebT)} + \frac{e^2}{6}(a^2 - b^2) - \frac{1}{T^2}\right),$$
$$a^2 - b^2 = B^2 - E^2,\ ab = \boldsymbol{E}\cdot\boldsymbol{B}$$

(1)

Then the integrals are expanded in the series [7]:

$$L = L_{spin} + 2L_{scal},$$
$$L_{spin} = -\frac{m^4}{8\pi^2}\sum_{N=4}^{\infty}\left(\frac{2e}{m^2}\right)^N \sum_{K=0}^{N} C_{spin}^{(1)}\left(\frac{K}{2}, \frac{N-K}{2}\right) X_+^{K/2} X_-^{(N-K)/2},$$
$$L_{scal} = -\frac{m^4}{16\pi^2}\sum_{N=4}^{\infty}\left(\frac{2e}{m^2}\right)^N \sum_{K=0}^{N} C_{scal}^{(1)}\left(\frac{K}{2}, \frac{N-K}{2}\right) X_+^{K/2} X_-^{(N-K)/2},$$
$$X_+ = \frac{\boldsymbol{B}^2 - \boldsymbol{E}^2}{4} + i\frac{(\boldsymbol{EB})}{2},\quad X_- = \frac{\boldsymbol{B}^2 - \boldsymbol{E}^2}{4} - i\frac{(\boldsymbol{EB})}{2}$$

(2)

Moreover, the coefficients $C_{spin}^{(1)}, C_{scal}^{(1)}$ are determined by the formulas [7]:

$$C_{spin}^{(1)}\left(\frac{K}{2}, \frac{N-K}{2}\right) = (-1)^{0.5N}(N-3)!\sum_{k=0}^{K}\sum_{l=0}^{N-K}(-1)^{N-K-l}\frac{B_{k+l}B_{N-k-l}}{k!\,l!(K-k)!(N-K-l)!},$$
$$C_{scal}^{(1)}\left(\frac{K}{2}, \frac{N-K}{2}\right) = (-1)^{0.5N}(N-3)!\sum_{k=0}^{K}\sum_{l=0}^{N-K}(-1)^{N-K-l}\frac{(1-2^{1-k-l})(1-2^{1-N+k+l})B_{k+l}B_{N-k-l}}{k!\,l!(K-k)!(N-K-l)!},$$

(3)

where $B_k$ are the Bernoulli numbers.

We consider six orders inclusive. Let us write out the values of nonzero coefficients $C_{spin}^{(1)}, C_{scal}^{(1)}$ (table 1).

Table 1.

Values of the spinor and scalar QED coefficients for one-loop photon amplitudes

| $(K,N)$ | $C^{(1)}_{spin}$ | $C^{(1)}_{scal}$ |
|---|---|---|
| $(0,4)$ | 4,167·10⁻³ | 4,167·10⁻³ |
| $(2,4)$ | -0,031 | 0,011 |
| $(4,4)$ | 4,167·10⁻³ | 4,167·10⁻³ |
| $(0,6)$ | 9,921·10⁻⁴ | 9,921·10⁻⁴ |
| $(2,6)$ | -7,341·10⁻³ | 5,159·10⁻³ |
| $(4,6)$ | -7,341·10⁻³ | 5,159·10⁻³ |
| $(6,6)$ | 9,921·10⁻⁴ | 9,921·10⁻⁴ |

The system of Maxwell equations (with explicitly expressed vectors **D** and **H** through the vectors **E** and **B**) can be written in the form:

$$\nabla \cdot \mathbf{E} = (\rho - \nabla \cdot \mathbf{P})/\varepsilon_0,$$
$$\nabla \cdot \mathbf{B} = 0,$$
$$\frac{\partial \mathbf{B}}{\partial t} + \nabla \times \mathbf{E} = 0, \tag{4}$$
$$\frac{1}{c^2}\frac{\partial \mathbf{E}}{\partial t} - \nabla \times \mathbf{E} = -\mu_0\left(\mathbf{j} + \frac{\partial \mathbf{P}}{\partial t} + \nabla \times \mathbf{M}\right)$$

This system, in the absence of free charges and currents, can be rewritten as [8]:

$$\frac{1}{c^2}\frac{\partial^2 \mathbf{E}}{\partial t^2} - \nabla^2 \mathbf{E} = -\mu_0\left(\frac{\partial^2 \mathbf{P}}{\partial t^2} + c^2\nabla(\nabla \cdot \mathbf{P}) + \frac{\partial}{\partial t}(\nabla \times \mathbf{M})\right),$$
$$\frac{1}{c^2}\frac{\partial^2 \mathbf{B}}{\partial t^2} - \nabla^2 \mathbf{B} = \mu_0\left(\nabla \times (\nabla \times \mathbf{M}) + \frac{\partial}{\partial t}(\nabla \times \mathbf{P})\right) \tag{5}$$

To establish a connection between the vectors **M** and **P** with the vectors **E** and **B** we used the Lagrangian (2). Taking into account that $\mathbf{P} = \delta L/\delta \mathbf{E}$, $\mathbf{M} = -\delta L/\delta \mathbf{B}$ and, using the equality of some coefficients (Table 1), we can obtain expressions for the magnetization and polarization:

$$P = \frac{m^4 e^4}{8\pi^2}\left(C^1_{spin}(2,4) - C^1_{scal}(2,4)\right)\left(\frac{\bm{B}}{2}(\bm{E}\cdot\bm{B}) - \frac{\bm{E}}{4}\left(c^2\bm{B}^2 - \bm{E}^2\right)\right) + \frac{m^4 e^6}{4\pi^2}\left(C^1_{spin}(2,6) - C^1_{scal}(2,6)\right)\cdot$$

$$\left(\frac{-\bm{E}}{2}\left(\left(\frac{c^2\bm{B}^2 - \bm{E}^2}{16}\right)^2 - \frac{(\bm{E}\cdot\bm{B})^2}{4}\right) + \frac{\bm{B}(c^2\bm{B}^2 - \bm{E}^2)(\bm{E}\cdot\bm{B})}{8} - \bm{E}\left(\frac{(c^2\bm{B}^2 - \bm{E}^2)^2}{16} + \frac{(\bm{E}\cdot\bm{B})^2}{4}\right)\right),$$

$$M = -\frac{m^4 e^4}{8\pi^2}\left(C^1_{spin}(2,4) - C^1_{scal}(2,4)\right)\left(\frac{\bm{E}}{2}(\bm{E}\cdot\bm{B}) + \frac{\bm{B}}{4}\left(c^2\bm{B}^2 - \bm{E}^2\right)\right) - \frac{m^4 e^6}{4\pi^2}\left(C^1_{spin}(2,6) - C^1_{scal}(2,6)\right)\cdot$$

$$\left(\frac{\bm{B}}{2}\left(\left(\frac{c^2\bm{B}^2 - \bm{E}^2}{16}\right)^2 - \frac{(\bm{E}\cdot\bm{B})^2}{4}\right) + \frac{\bm{E}(c^2\bm{B}^2 - \bm{E}^2)(\bm{E}\cdot\bm{B})}{8} - \bm{B}\left(\frac{(c^2\bm{B}^2 - \bm{E}^2)^2}{16} + \frac{(\bm{E}\cdot\bm{B})^2}{4}\right)\right)$$

(6)

### 3. Main results of numerical simulation

We choose the initial conditions for the system of equations (5.6) in the form:

$$E_z = 0,$$
$$B_x = B_y = B_z = 0,$$
$$E_x = E_y = A \cdot exp\left(-\left(\frac{z}{\gamma}\right)^2\right) exp\left(-\frac{x^2 + y^2}{\gamma_p^2}\right),$$
$$\frac{d}{dt}E_z = \frac{d}{dt}B_x = \frac{d}{dt}B_y = \frac{d}{dt}B_z = 0,$$
$$\frac{d}{dt}E_x = \frac{d}{dt}E_y = \frac{2vA}{\gamma^2} \cdot exp\left(-\left(\frac{z}{\gamma}\right)^2\right) exp\left(-\frac{x^2 + y^2}{\gamma_p^2}\right)$$

(7)

which corresponds to the propagation at the initial instant of a cylindrically symmetric pulse of a Gaussian electric field. In Eq. (7) $v$ is the initial velocity of the pulse, $\gamma$ is the width of the pulse along the direction of propagation, $\gamma_p$ is the width of the pulse in the direction perpendicular to the propagation direction. Subsequently, we use the cylindrical coordinate system:

$$\Delta \bar{B} = \left(\Delta B_\rho - \frac{B_\rho}{\rho^2}\right)\hat{\rho} + \left(\Delta B_\varphi - \frac{B_\varphi}{\rho^2}\right)\hat{\varphi} + \Delta B_z \cdot \hat{z},$$

$$\bar{\nabla} \times \bar{B} = -\frac{\partial B_\varphi}{\partial z}\hat{\rho} + \left(\frac{\partial B_\rho}{\partial z} - \frac{\partial B_z}{\partial \rho}\right)\hat{\varphi} + \frac{1}{\rho}\frac{\partial(\rho B_\varphi)}{\partial \rho}\hat{z},$$

$$\bar{\nabla} \cdot \bar{B} = \frac{1}{\rho}\frac{\partial(\rho B_\rho)}{\partial \rho} + \frac{\partial B_z}{\partial z},$$

$$\Delta f = \frac{1}{\rho}\left(\frac{\partial}{\partial \rho}\left(\rho \frac{\partial f}{\partial \rho}\right)\right) + \frac{\partial^2 f}{\partial z^2},$$

$$\bar{\nabla} f = \frac{\partial f}{\partial \rho}\cdot \hat{\rho} + \frac{\partial f}{\partial z}\cdot \hat{z},$$

$$\rho = \sqrt{x^2 + y^2},\ tg\varphi = \frac{y}{x}$$

(8)

and apply a numerical cross-type scheme [9].

The typical evolution of the momentum for Lagrangian (2) is shown in Fig. 1.

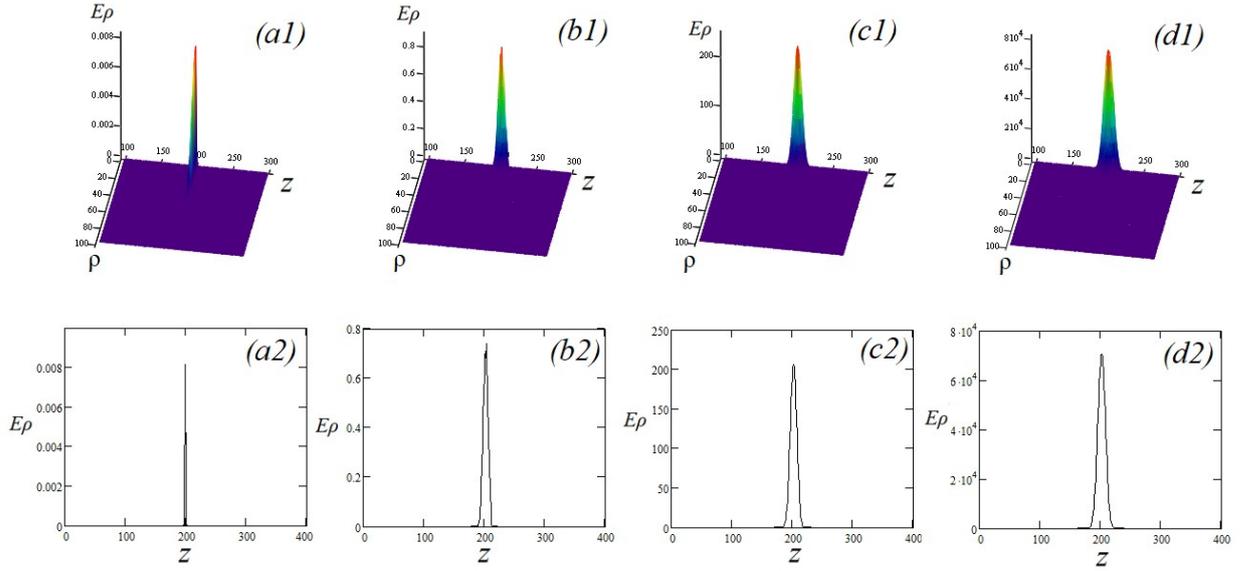

Fig. 1. Dependence of the component of the electric field strength $E_\rho$ on the coordinates at different times: (*a*) $t=0$; (*b*) $t=150$; (*c*) $t=250$; (*d*) $t=350$. A three-dimensional view - figures with index 1 and longitudinal sections below it for each moment in time - figures with index 2. All values are in relative units.

The pulse collapse is observed, which can be seen earlier in the framework of nonlinear electrodynamics in other works [8, 10–12].

Note that for the given initial conditions (7), the electric field $E$ has the greatest value at all times. This is shown in table 2 and table 3. In this regard, we restrict ourselves to considering only the field *E*, and we use the Schwinger formula, but not by the more complete formula given in [13].

Table 2.

Maximum values of ***E***$^2$ and ***B***$^2$ with nonzero initial conditions for *E*.

| t | 0 | 50 | 100 | 150 | 200 | 250 | 300 | 350 |
|---|---|---|---|---|---|---|---|---|
| max(***E***$^2$) | 0,008173 | 0,018 | 0,066 | 0,888 | 13,89 | 246,4 | 4529 | 81030 |
| max(***B***$^2$) | 0 | 0 | 0 | 0 | 0 | 2,64E-14 | 4,33E-08 | 0,086 |

Table 3.

Maximum values of ***E***$^2$ and ***B***$^2$ with nonzero initial conditions for *B*.

| t | 0 | 50 | 100 | 150 | 200 | 250 | 300 | 350 |
|---|---|---|---|---|---|---|---|---|
| max(***E***$^2$) | 0 | 0 | 0 | 0 | 0 | 0 | 0 | 0 |
| max(***B***$^2$) | 0,008162 | 0,006144 | 0,002451 | 0,003086 | 0,002808 | 0,002444 | 0,002104 | 0,001821 |

As well-known, in the collapse zone the Schwinger mechanism of pair production begins to work. The velocity of formation of electron-positron pairs according to the Schwinger mechanism [3] for scalar (*j*=0) and spin (*j*=0.5) particles can be determined by the formulas:

$$w = \frac{(2j+1)}{8\pi^3} \sum_{k=1}^{\infty} (-1)^{(2j+1)(k+1)} \left(\frac{eE}{k}\right)^2 \cdot e^{-\frac{k \cdot E_{cr}}{|E|}},$$

$$E_{cr} = \frac{\pi m^2 c^3}{\hbar e}$$

(9)

Due to the Schwinger mechanism, as soon as the field becomes larger than the critical one, intensive pair production begins, which is obviously easy to verify by experimental methods. Note that Schwinger's formula is applicable to homogeneous and static fields. In this paper, it is applied to inhomogeneous and variable fields to estimate the velocity of pair creation. Naturally, it is necessary to take into account the effects of inhomogeneity and variability of the electric field, but it is precisely to estimate the velocity of pair creation we use the Schwinger formula.

The velocity of the creation of electron-positron pairs depending on the coordinates is shown in Fig. 2.

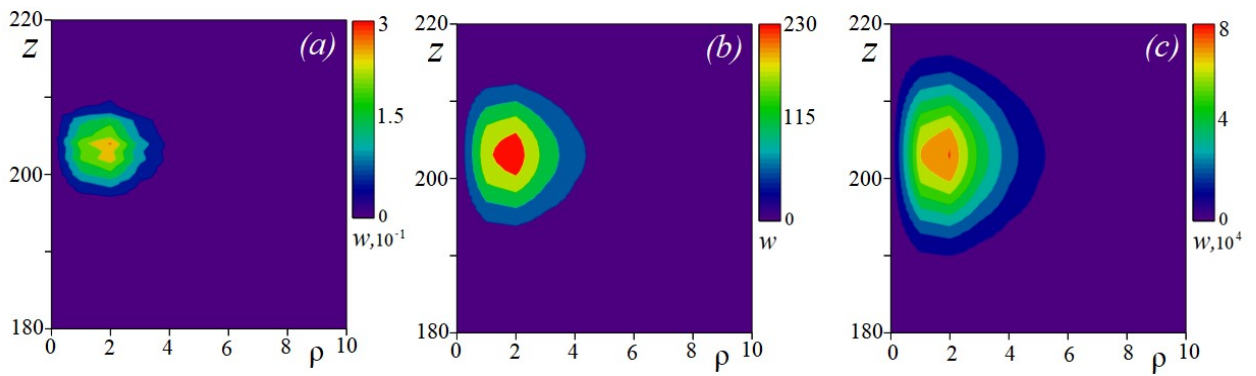

Fig. 2. The velocity of the electron-positron pairs creation in the case of scalar particles for different times (level lines): (a) *t*=150; (b) *t*=250; (c) *t*=350. All values are in relative units.

Note that in the case of spin particles, the velocity of the particle creation has a similar character.

**4. Conclusion**

The dependence of the creation velocity on the coordinates shows that there are two cylindrical regions in which pairs are produced. This effect can be experimentally detected.

It should be noted that the pairs creation leads to a decrease in the pulse energy. In addition, there is another very important effect, namely, the creation of pairs leads to the formation of an electron-positron plasma. Due to this plasma has finite conductivity, further pulse absorption occurs in it. This leads to the self-consistent problem of the propagation of an electromagnetic pulse in an electron-positron plasma. In this case, additional terms in Maxwell's equations of the form (8) should also be taken into account. Obviously, this represents a separate problem, which is planned to be considered in the future.

Let's formulate the main conclusions from the work:

1) The solution of Maxwell's equations for supersymmetric nonlinear electrodynamics is obtained, which demonstrates the collapse of the pulse.

2) The scenario of the onset of the pulse collapse is described in the framework of supersymmetric nonlinear electrodynamics. This scenario includes a sharp increase in the pulse in the direction of propagation for a finite period of time.

3) The velocity of the creation electron-positron pairs by the Schwinger mechanism is estimated. It is shown that pairs are produced in two cylindrical symmetric regions in the form of a ring (due to the initial pulse symmetry), which can be found experimentally.

**5. Acknowledgment**

Authors thanks the Ministry of Science and Higher Education of the Russian Federation for the numerical modeling and parallel computations support under the government task (0633-2020-0003). Also Belonenko M.B. thanks the Russian Foundation for Basic Research and the Volgograd Region Administration for support within the framework of the scientific project No. 19-43-340005 r_a.